\documentclass[prb,twocolumn,preprintnumbers,amsmath,amssymb,floatfix,superscriptaddress]{revtex4-1}
\usepackage{graphicx}
\usepackage{epstopdf}
\usepackage{dcolumn}
\usepackage{bm}
\usepackage{color}
\usepackage{natbib}
\usepackage{amsmath}
\usepackage{amssymb}
\usepackage{multirow} 

\begin{document}

\title{Turning ABO$_3$ antiferroelectrics into ferroelectrics: Design rules for practical rotation-driven ferroelectricity in double perovskites and A$_3$B$_2$O$_7$ Ruddlesden-Popper }

\author{Andrew T. Mulder}
\affiliation{School of Applied \& Engineering Physics, Cornell University, Ithaca, New York 14853, USA}
\author{Nicole A. Benedek}
\affiliation{Materials Science and Engineering Program, The University of Texas at
Austin, Austin, Texas 78712, USA}
\author{James M.\ Rondinelli}
\affiliation{Department of Materials Science \& Engineering,\!
	Drexel University,\! Philadelphia,\! PA 19104,\! USA}%
\author{Craig J. Fennie}
	\email{fennie@cornell.edu}
\affiliation{School of Applied \& Engineering Physics, Cornell University, Ithaca, New York 14853, USA}

\maketitle

\section{abstract}

Ferroic transition metal oxides, which exhibit spontaneous elastic, electrical, magnetic or toroidal order, exhibit functional properties that find use in ultrastable solid-state memories to sensors and medical imaging technologies. To realize multifunctional behavior, where one order parameter can be coupled to the conjugate field of another order parameter, however, requires a common microscopic origin for the long-range order. Here, we formulate a complete theory for a novel form of ferroelectricity, whereby a spontaneous and switchable polarization emerges from the destruction of an antiferroelectric state due to octahedral rotations and ordered cation sublattices. We then construct a materials design framework based on crystal-chemistry descriptors rooted in group theory, which enables the facile design of artificial oxides with large electric polarizations, $P$, simultaneous with small energetic switching barriers between +$P$ and -$P$. We validate the theory with first principles density functional calculations on more than 16 perovskite-structured oxides, illustrating it could be operative in any materials classes exhibiting two- or three-dimensional corner-connected octahedral frameworks. We show the principles governing materials selection of the ``layered'' systems originate in the lattice dynamics of the A cation displacements stabilized by the pervasive BO$_6$ rotations of single phase ABO$_3$ materials, whereby the latter distortions govern the optical band gaps, magnetic order and critical transition temperatures. Our approach provides the elusive route to the ultimate multifunctionality property control by an external electric field.


\section{Introduction}


In the search for new classes of multifunctional materials, the design or discovery of ferroelectrics in which the spontaneous electrical polarization couples strongly to other structural, magnetic, orbital, and electronic degrees of freedom is a challenge being actively pursued as a means to achieve electric field-controllable emergent phenomena such as ferromagnetism~\cite{martin12,he12}.
Much of the current materials-by-design effort has focused on the structurally and chemically complex ABO$_3$ perovskites, a large class of functional materials that display a wide range of properties due to their highly tunable ground states.
Because of the high susceptibility of perovskite materials towards polar structural instabilities, a notion has emerged that it is generally more productive to start with a material that displays, for example, ferromagnetism, and devise a way to induce  ferroelectricity.  Two highly successful approaches that have captivated the attention of researchers over the last decade are that of epitaxial strain engineering (strain-induced ferroelectricity, mutliferroicity) and that of selective chemical substitution of a stereochemically inactive cation with a lone-pair-active cation such as Bi$^{3+}$, as in BiFeO$_3$~\cite{hill00,belik12}. 
While highly successful at creating new multiferroics (materials that are both ferromagnetic and ferroelectric), generally speaking these approaches have not  led to a widespread solution to the central problem  of {\it strong coupling} between the polarization and the magnetism. The reason is believed to be due to the common nature of the ferroelectricity in these materials (small cation displacements such as those in the prototypical perovskite ferroelectric BaTiO$_3$). New ideas to realize ferroelectricity are clearly needed.

\begin{figure}[t]
\centering
\includegraphics[width=9.0cm]{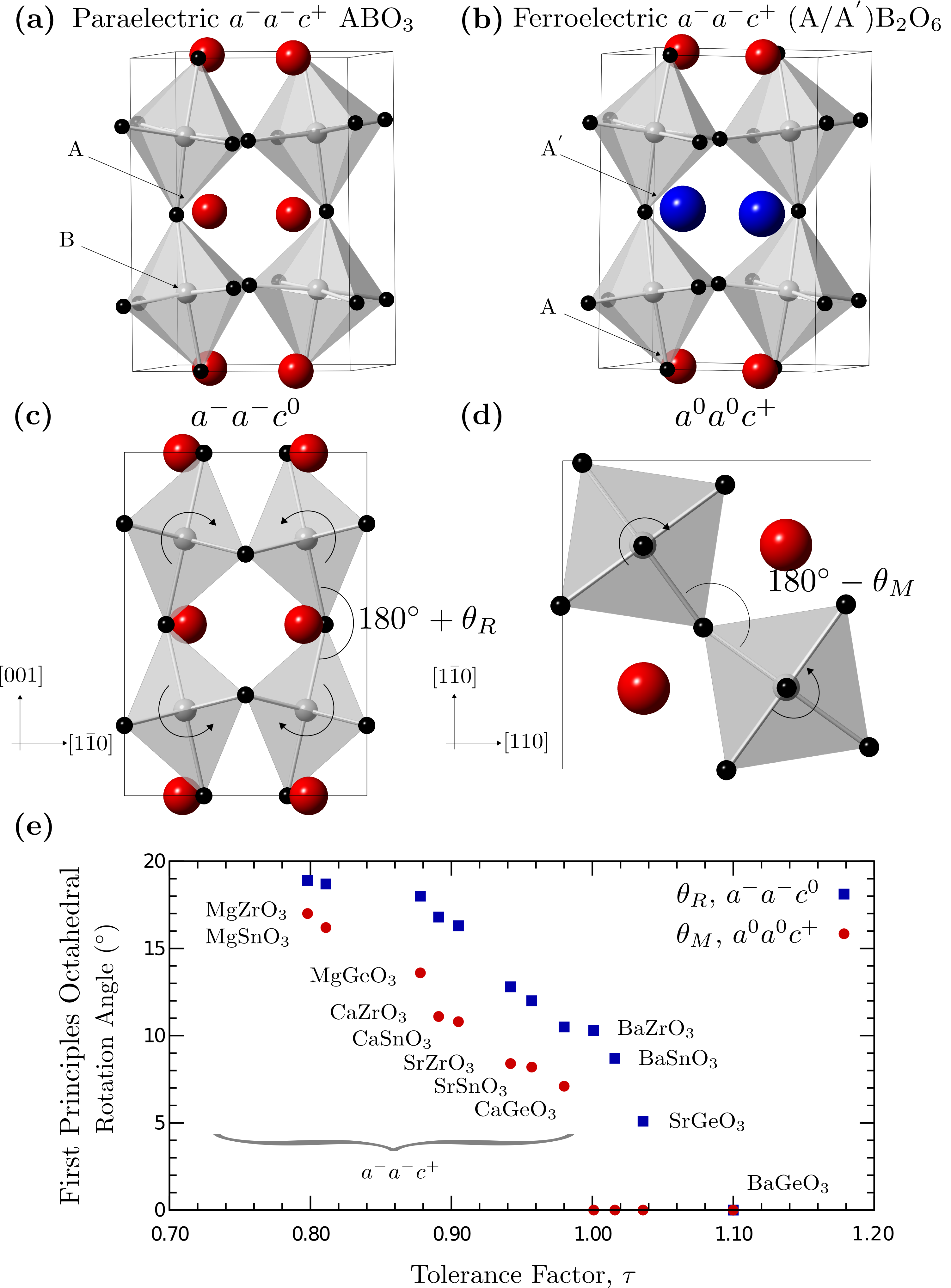}\vspace{-0pt}
\caption{(a) The paralectric \textit{Pnma} structure (Glazer rotation pattern a$^-$a$^-$c$^+$) of the ABO$_3$ constituents, (b) the ferroelectric (A/A$^{'}$)B$_2$O$_6$ structure with the same a$^-$a$^-$c$^+$ rotation pattern, (c) the a$^-$a$^-$c$^0$ and (d) the a$^0$a$^0$c$^+$ rotation patterns that make up \textit{Pnma}. (e) First principles calculated octahedral rotations of a suite of materials that span a wide range of tolerance factor.}
\label{Fig1}
\end{figure}

Although perovskites are often what comes to mind when discussing oxide ferroelectricity, the overwhelming majority of oxide perovskites -- particularly those which have active electronic, magnetic, and orbital degrees of freedom --  adopt highly distorted, non-polar, ground state structures in which the BO$_6$ octahedra are rotated about one or more of the crystal axes~\cite{lufaso01}. A fascinating question that has only recently been considered in earnest, starting with the work of Ref.~\onlinecite{bousquet08}, concerns how to directly control these  octahedral rotations with an external electric field. Since rotations of the BO$_6$ octahedra couple strongly to other properties~\cite{kimura03b,junquera11,rondinelli12,caviglia12}, magnetism~\cite{goto04} for example, gaining control over them could solve the strong coupling problem~\cite{benedek12}.  Another way of stating the challenge is, ``how can octahedral rotations induce a spontaneous polarization?''

By themselves octahedral rotations cannot induce ferroelectricity in simple perovskites, but recent work has demonstrated that they can induce ferroelectricity in layered A-site ordered double perovskites and superlatties~\cite{knapp06,bousquet08,king10,rondinelli11,fukushima11} and in Ruddlesden-Popper phases~\cite{aleksandrov95,benedek11} (and  other layered perovskites~\cite{levin99,ederer06,jorge11}). 
The most common realization of this novel rotation-centric, ferroelectric mechanism has been referred to as hybrid improper ferroelectricity (HIF)~\cite{ benedek11}. 
The defining feature of HIF is a symmetry-allowed trilinear coupling in the free energy, 
\begin{equation}
\mathcal{F}_{\rm tri} = \gamma Q_{R1} Q_{R2} P, 
\label{tri}
\end{equation}
where $P$ is the amplitude of the polarization,  $Q_{R1}$ and $Q_{R2}$ are the amplitudes of non-polar, symmetry inequivalent octahedral rotation modes~\cite{ghosez11}, and $\gamma$ is the coupling coefficient. The presence of this  invariant in the free energy implies that when $Q_{R1}$ and $Q_{R2}$ become non-zero, a polarization will be induced in the ground state structure, even in the absence of prototypical, BaTiO$_3$-like, polar (zone-center) instabilities. 
%


\section{Simple Heuristic Design Rules from first-principles data}


\subsection{Recent work on rotation driven ferroelectrics by design and a problem}

Using a combination of first-principles electronic structure methods and symmetry arguments, Rondinelli and Fennie recently established design rules for the creation of hybrid improper ferroelectrics in 
ABO$_3$/A$^\prime$BO$_3$  superlattices~\cite{rondinelli11}. The aforementioned superlattices correspond to  (A/A$^\prime$)B$_2$O$_6$ double perovskites in which the A and A$^\prime$ sites  order into alternating layers along [001] (note we will use these two different views of the structure interchangeably). 
These rules can be summarized as follows:  (1) the chemical criterion states that ABO$_3$/A$^\prime$BO$_3$ superlattices allows for a trilinear invariant, $\mathcal{F}_{\rm tri}$, in the free energy by symmetry, and (2) the energetic criterion states that at least one of the perovskite constituents of the superlattice should have a strong tendency towards the $Pnma$ perovskite structure (that is, $Pnma$ should be the ground state structure, preferably, or a metastable phase with a wide stability window). 
The majority of perovskites form in the \textit{Pnma} space group~\cite{woodward97a,woodward97b,lufaso01} and therefore the rules of Rondinelli and Fennie are widely accessible to many chemistries and have the potential to lead to new classes of  multifunctional materials. Note that the symmetry of the \textit{Pnma} structure is established by two symmetry-inequivalent octahedral rotations, which in Glazer notation  are: a$^-$a$^-$c$^0$ (with amplitude $Q_R$ because this rotation pattern  transforms like the irreducible representation (irrep)  $R_4^+$  of $Pm\bar{3}m$) and a$^0$a$^0$c$^+$ (with amplitude $Q_M$, which transforms like $M_3^+$), shown in Figures~\ref{Fig1} (c) and (d) respectively.  The energetic criterion ensures that the combined rotation pattern a$^-$a$^-$c$^+$ survives in the ordered double perovskite.

These design rules establish when such a hybrid improper state should exist. The microscopic mechanism responsible for the polarization, however, is still unknown and hence the design rules do not directly address the question of whether or not this  hybrid improper state is a functional ferroelectric or  simply a pyroelectric. Without insight from the microscopics,  a  fundamental materials problem that  prevents widespread realization of these new multifunctional materials remains -- that of understanding how to design a material with a large spontaneous polarization and a low ferroelectric switching barrier.  This problem is best described by considering:
\begin{itemize}
\item[]  \textbf{Conjecture 1.} Given the form of the trilinear coupling term,  Eq.~\ref{tri}, it is  reasonable to assume that the spontaneous polarization $P$ will increase as the strength of  each rotation, $Q_{R1}$ and $Q_{R2}$, increases. That is, as the energetic criterion becomes increasingly satisfied. Indeed, to lowest order the  polarization can be shown to be proportional to
\begin{equation}
P = \gamma {Q_{R1} Q_{R2}\over \tilde{A}_P}
\label{Ptri}
\end{equation}
where $\tilde{A}_P$ is the polar mode stiffness renormalized by rotations and other structural distortions  (see Appendix for details). 

\item[] \textbf{Corollary 1.} If this assumption is true there is a problem, in that large rotations necessarily lead to a large ferroelectric switching barrier (barring pathologically flat energy surfaces), since in order to switch   $P$ from say up to down, you must switch the sense of one of the BO$_6$ octahedral rotations.
\end{itemize}
Are these two points really true, and if so, is there a way around them in order to realize a high $P$, low switching barrier rotation-driven ferroelectric? 
%
 
\begin{figure*}[t]
\centering
\includegraphics[width=17.0cm]{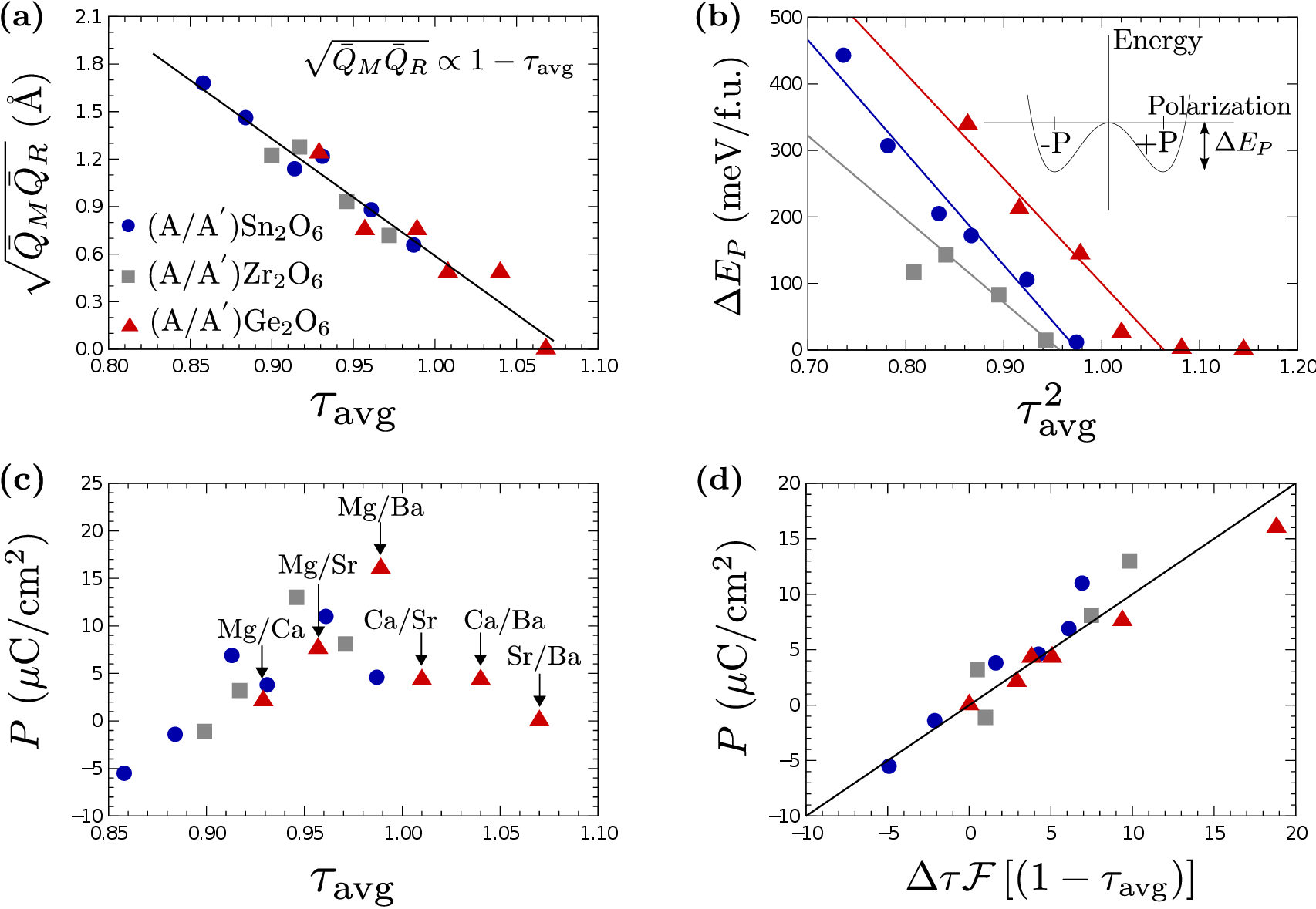}\vspace{-0pt}\\
\caption{ABO$_3$/A$'$BO$_3$ superlattices: (a) Expected strength of a$^-$a$^-$c$^+$ rotations based on properties of ABO$_3$ constituents,  (b) Stabilization energy (Note: in a proper ferroelectric transition with an order parameter Q, $\Delta E \propto Q$. Since in our model $Q \propto \tau_{\rm avg}^2$ we plot $\Delta E$ as linear in $\tau_{\rm avg}^2$), (c) Polarization versus average tolerance factor with stannates highlighted, and (d) Polarization versus the tolerance factor renormalized by the tendency for A-site displacements showing an almost perfect fit to the first principles data.}
\label{tolvs}
\end{figure*}

  \subsection{Materials Suite to Test Conjecture}

One way to control the magnitude of octahedral rotations in perovskites is by chemical substitution of A or B-site cations with different ionic radii, $r_A$ and/or $r_B$. This effectively allows one to control the tolerance factor,  
\begin{equation}
\tau\equiv { r_A+r_O \over   \sqrt{2}(r_B+r_O) },  
\end{equation}
a geometric descriptor that correlates with the stability of a particular ABO$_3$ material towards octahedral rotations~\cite{lufaso01}. In general, as $\tau$ decreases, the greater the susceptibility of a material towards octahedral rotations.  
We consider three families of perovskites and within each family we consider A = Mg, Ca, Sr, and Ba, which as shown in Figure~\ref{Fig1}(e),  span a wide range of tolerance factors and therefore have a wide favorability towards $Pnma$. All exist in nature in the perovskite structure except MgZrO$_3$ and MgSnO$_3$.

In Figure~\ref{tolvs}(a) we plot a measure of the expected magnitude of the a$^-$a$^-$c$^+$ rotation pattern in the A/A$'$ double perovskites versus average tolerance factor, $\tau_{\rm avg}$ (defined as the average tolerance factor of ABO$_3$ and A$^\prime$BO$_3$).  This measure, based on the properties of the $Pnma$ perovskites, is the geometric mean of $\bar{Q}_{M}$ and $\bar{Q}_{R}$, where both quantities are defined as,
\begin{equation}
\bar{Q}_{M} =  \frac{1}{2} \left(\langle Q_{M}^{ABO_3}\rangle+ \langle Q_{M}^{A'BO_3}\rangle\right),
\end{equation}
where $\langle Q_{M}^{ABO_3}\rangle$  and $\langle Q_{M}^{A^\prime BO_3}\rangle$ are the amplitudes of $M_3^+$ in $Pnma$ ABO$_3$  and A$^\prime$BO$_3$ respectively.  A nearly linear behavior can be seen, and as expected, the smaller the average tolerance factor, the larger the magnitude of the rotations.~\cite{lufaso01}

In Figure~\ref{tolvs}(b) we plot the energy difference, $\Delta$E$_{\rm P}$, between the lowest energy paraelectric state (the fully relaxed a$^-$a$^-$c$^0$ structure in space group $Pmma$), and the polar state in the actual ABO$_3$/A$'$BO$_3$ system.   By using this paraelectric state as a reference structure this not only gives us an idea as to the `stability' of the hybrid improper ground state but also the intrinsic energetic barrier to switch the polarization -- via switching the $a^0a^0c^+$ rotation -- between ferroelectric domains.  
Across our entire test suite of materials of sixteen different superlattices (SLs), we find that the stability of the hybrid improper state smoothly increases as the average tolerance factor decreases, consistent with the energetic criterion of Rondinelli and Fennie, which subsequently  implies that the  intrinsic barrier to switch the polarization also increases. 
Therefore, in order to have a ferroelectric that is switchable in an experimentally realizable electric field the  average tolerance factor of the perovskite constituents making up the superlattice will need to be large. {\it Does this mean that the  spontaneous polarization will necessarily be small,   as alluded to by  Corollary 1?}

We now consider how the induced polarization varies with $\tau_{\rm avg}$. In Figure~\ref{tolvs}(c) we plot the calculated polarization of the various superlattices versus average tolerance factor where it is seen that no obvious trend  exists across all sixteen SLs. In fact, it appears to be maximal near $\sim$0.98.
The lack of a clear correlation between $P$ and the magnitude of the rotations  is in stark contrast to the expectation based on the trilinear coupling picture. Indeed from Equation~\ref{Ptri} and our first-principle results, which indicate that $ \bar{Q}_M\bar{Q}_R\propto (1-\tau_{\rm avg})^2$ as shown in Figure~\ref{tolvs}(a) and that   $\tilde{A}_P $ is largely independent of tolerance factor (as expected  for the renormalized value in $Pnma$ as shown in Fig.~S\ref{supp_polarmode}), the trilinear coupling picture would predict 
\begin{equation}
 P_{\rm tri} \propto {\bar{Q}_{M} \bar{Q}_{R}\over \tilde{A}_P}  \propto (1- \tau_{\rm avg})^2,
\label{PtriNOT}
\end{equation}
where the renormalized force constant of the polarization, $\tilde{A}_P$, is independent of $\tau$.
%
\footnote{Note, this can be understood by realizing that  $\tilde{A}_P$ is essentially the average polar force constant of ABO$_3$ and A$'$BO$_3$ in the Pnma structure and that $\tilde{A}_P$ in each Pnma compound is independent of $\tau$. See Appendix Figure~\ref{supp_polarmode}.} 
Figure~\ref{tolvs}(c), however, makes clear that there is not this (or any) simple, universal, chemistry-independent correlation between the polarization and the magnitude of the rotations:
\begin{equation}
P \not\propto (1-\tau_{\rm avg})^2.
\label{PNOT}
\end{equation}
What went wrong, if anything, in this argument?  Perhaps the expected magnitude of the rotations in the superlattices is not approximated well by the averages? While this may in fact lead to a small discrepancy, it is unlikely that this is the origin of the qualitative disagreement between the expected polarization within the trilinear coupling picture and one's own chemical and physical intuition.   Instead, the key is to realize that the coupling coefficient, $\gamma$, can have a nontrivial, non-monotonic dependence on the particular ABO$_3$ materials making up the superlattice. In order to regain an intuitive and predictive relationship between the polarization and simple chemical descriptors such as a tolerance factor, however, the microscopic origin of  the coupling constant needs to be addressed. Does $\gamma$ really depend on the specific material constituents of the superlattice or can it be described by a universal functional form that is independent of chemistry?

To address this question, let us first look more closely at the polarization within a specific family of compounds, the germanate, AGeO$_3$/A$'$GeO$_3$ SLs.  As shown in Figure~\ref{tolvs}(c), the polarization initially increases as $\tau$ decreases, but then remains fairly constant as $\tau_{\rm avg}$ is  further reduced, but then increases substantially before steadily decreasing as average tolerance factor is still further reduced. As already concluded, this is not the  $\tau_{\rm avg}$ dependence expected from Equations~\ref{PtriNOT} and~\ref{PNOT}.

If, however, a similar exercise is performed but this time keeping the chemistry of one of the A-sites  fixed, a pattern begins to emerge. This is best seen by considering two cases. First, consider the AGeO$_3$/BaGeO$_3$ SLs as the average tolerance factor decreases:
\begin{itemize}
\item[(1)]  SrGeO$_3$/BaGeO$_3$, $\tau_{\rm avg}$ = 1.07, P=0$\mu$C/cm$^2$  \\
 \textcolor{white}{q}$\searrow $\\ 
 \textcolor{white}{q}  CaGeO$_3$/BaGeO$_3$, $\tau_{\rm avg}$ = 1.04, P=4$\mu$C/cm$^2$  \\
  \textcolor{white}{qqq}$\searrow$\\ 
\textcolor{white}{qqq}  MgGeO$_3$/BaGeO$_3$, $\tau_{\rm avg}$ = 0.99, P=16$\mu$C/cm$^2.$ 
\end{itemize}
Here  the polarization monotonically increases, consistent with one's intuition. Next consider the MgGeO$_3$/A$'$GeO$_3$ SLs:
\begin{itemize}
\item[(2)]   MgGeO$_3$/BaGeO$_3$, $\tau_{\rm avg}$ = 0.99, P = 16$\mu$C/cm$^2$  \\
 \textcolor{white}{qqqqqqqqqq}$\searrow$\\ 
 \textcolor{white}{qq}  MgGeO$_3$/SrGeO$_3$, $\tau_{\rm avg}$ = 0.96, P = 7$\mu$C/cm$^2$ \\
  \textcolor{white}{qqqqqqqqqqqq}$\searrow$\\ 
\textcolor{white}{qqqq}  MgGeO$_3$/CaGeO$_3$, $\tau_{\rm avg}$ = 0.93, P = 2$\mu$C/cm$^2$
\end{itemize}
where even though the average tolerance factor is still decreasing, in contrast to case (1)  the polarization steadily {\it decreases}. What do we learn from this?

\begin{figure*}[t]
\centering
\includegraphics[height=4.5cm]{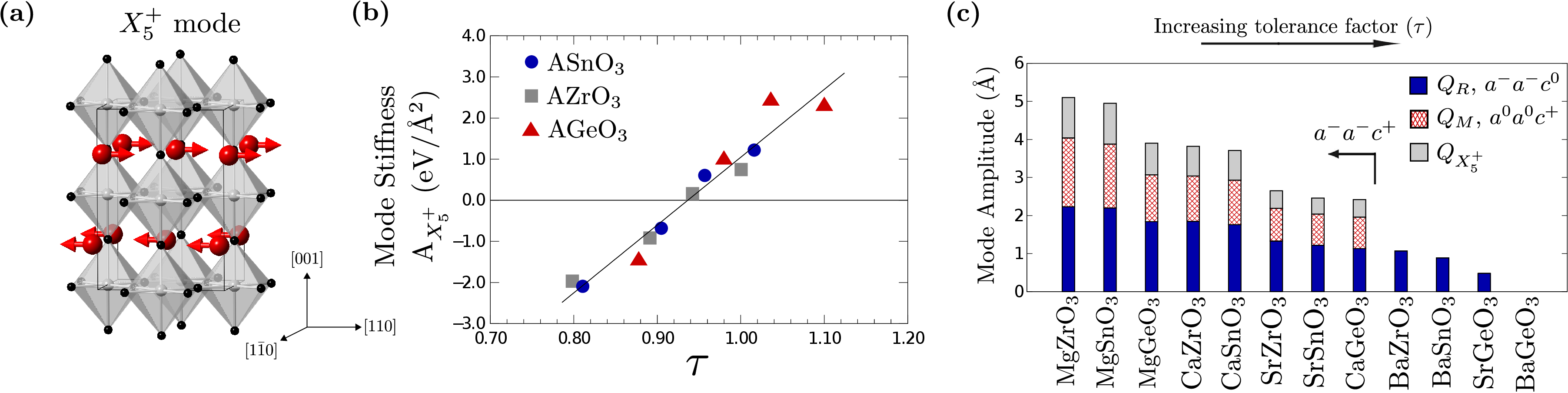}
\caption{(a) X$^+_5$ mode, (b) for each ABO$_3$ constituent: the stiffness of the antipolar $Q_{X_5^+}$ mode of Pm$\bar{3}$m. The ABO$_3$ cubic perovskite structure is unstable to the distortion when $A < 0$. (c) Amplitudes of the modes in the first-principles \textit{Pnma} structure of ABO$_3$ materials arranged by tolerance factor. In the \textit{Pnma} structure the octahedral rotations induce A-site displacements that transform like the irrep $X_5^+$ in $Pm\bar{3}m$.}
\label{Fig_X5}
\end{figure*}

Notice that in the first case the largest alkaline-earth cation we considered (Barium) is keep constant while the A-site cation steadily become smaller. In the second case, however, the smallest alkaline-earth cation we considered (Magnesium), is keep constant while the size of the A-cation is again steadily decreased. It would appear, at least within a family of compounds (for which the B-cation is keep constant), that the polarization is dominated by the difference in the ionic radii of the A and A$'$-cations.
By  elucidating the microscopic mechanism that leads to the trilinear coupling we will indeed argue that
\begin{equation}
\gamma \propto \Delta\tau \left(\gamma_1 - {\gamma_2\over 1-\tau_{\rm avg}} \right)
\end{equation}
where $\Delta\tau$ is the difference in tolerance factors of the ABO$_3$ constituents~\cite{rodriguez96,sinclair99}, while $\gamma_1$ and $\gamma_2$ are coupling coefficients originating from the fact that there are actually two distinct contributions to the polarization. This leads to a simple, chemically intuitive, universal result for the polarization: 
\begin{equation}
P \propto\Delta\tau \mathcal{F}\left[(1-\tau_{\rm avg})\right] \approx\Delta\tau (1-\tau_{\rm avg})
\end{equation}
where $\mathcal{F}$ is a smooth function of $(1-\tau_{\rm avg})$ and is approximately linear for all systems except those with very small tolerance factors (so small that either the perovskite phase doesn't form or the barrier to switching is so large that they are no longer good candidates for functional materials) for which the quadratic term must be considered. 

Now a clear correlation between the expected polarization and simple chemical descriptors -- $\tau_{\rm avg}$ and $\Delta\tau$ -- is realized as shown in Figure~\ref{tolvs}(d) where a nearly perfect fit over the entire tolerance factor range is seen.
This result implies that it is possible to simultaneously decrease the ferroelectric switching barrier, by increasing $\tau_{\rm avg}$, and increase the spontaneous polarization by increasing the difference in tolerance factors. Why? To answer this question we need to understand the origin of this rotation-induced ferroelectricity.


\section{Theory of the proposed Design Rules}


\subsection{Rotation-driven antiferroelectricity in Pnma perovskite}

Turning our attention back to the ABO$_3$ perovskite for a moment, it is well-known that rotations in perovskites are driven by the coordination preferences of the A-site cation~\cite{woodward97a,woodward97b}. The combination of octahedral rotation modes that establish the symmetry of the $Pnma$ structure -- $M_3^+$ and $R_4^+$ -- also favor displacements of the A-site cations. It turns out, because of the three dimensional  connectivity of the perovskite lattice, the A-sites displace in an antiferroelectric pattern, the exact motion of which can be thought of as local, polar displacements confined to the two-dimensional AO layers but arranged 180$^{\circ}$ out of phase along $\hat{z}$, as shown in Figure~\ref{Fig_X5}(a).  We denote the amplitude of this distortion as $Q_{X^+_5}$ since it transforms like the irrep $X_5^+$ in $Pm\bar{3}m$. When we decompose the fully relaxed $Pnma$ structures into symmetry-adapted basis functions of $Pm\bar{3}m$, we find that $Q_{X^+_5}$ is non-zero only when both $Q_M$ and $Q_R$ are non-zero, as shown in Figure~\ref{Fig_X5}(c). The $X_5^+$ mode itself, however, need not be unstable to appear in the $Pnma$ structure. Indeed, except for compounds with very small (for perovskites) tolerance factors ($\tau \lesssim $ 0.91), the test suite of materials considered do not favor this motion in the absence of the a$^-$a$^-$c$^+$ type of rotations, as indicated by the positive value of the force constant, $A_{X_5^+}$; see Figure~\ref{Fig_X5}(b). The appearance of a finite $Q_{X^+_5}$ in the $Pnma$ structure can be accounted for phenomenologically by a trilinear coupling in the free energy of $Pm\bar{3}$m,
\begin{equation}
\mathcal{F}_{\rm MRX^+_5} = \beta Q_M Q_R Q_{X^+_5}.
\label{eq2}
\end{equation}
This  implies that once the rotations become non-zero, a finite antiferroelectric structural distortion, $Q_{X^+_5}$, is induced~\cite{amisi12},
\begin{equation}
 Q_{X^+_5} \propto { Q_M  Q_R \over A_{X_5^+}}.
\end{equation}
The similarity between the trilinear coupling of rotations and the antiferroelectric distortion in the ABO$_3$ perovskites and the  trilinear coupling of rotations and the polarization the ABO$_3$/A$'$BO$_3$ perovskite SLs is curious. Could hybrid improper ferroelectricity actually originate from  antiferroelectricity?

\begin{figure}[t]
\centering
\includegraphics[width=8.0cm]{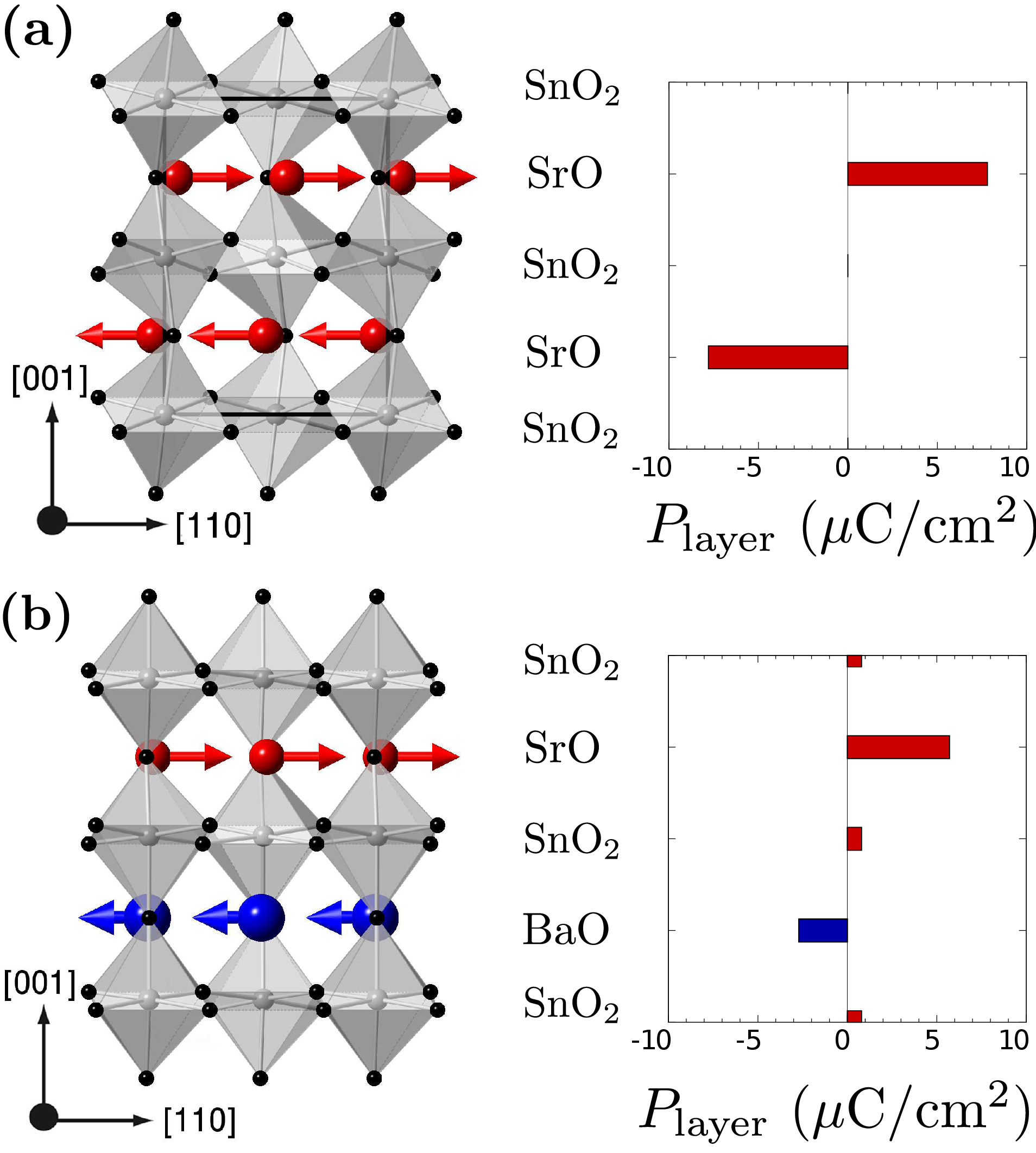}\vspace{-8pt}
\caption{(a)  \textit{Pnma} SrSnO$_3$ and the contributions to the polarization, $P = \sum P_{layer} = 0$, by symmetry. (b) (Sr/Ba)Sn$_2$O$_6$ superlattice where cancellation of $P_{layer}$ is not exact.}
\label{P_by_layer}
\end{figure}

In Figure~\ref{P_by_layer}(a) we plot a linear approximation to the layer-resolved polarization calculated from first principles (see Ref.~\onlinecite{wu06} for an exact method) for $Pnma$ SrSnO$_3$~\cite{green00} where the antiferroelectric order  can clearly be seen. With this  picture in mind it is now clear that replacing alternating AO layers with A$^\prime$O -- for example, replacing alternating SrO layers in SrSnO$_3$ with BaO to form (Sr/Ba)Sn$_2$O$_6$ --  creates chemically inequivalent A-sites. The small noncancellation of the layered polarization induced by the nominally antipolar $X_5^+$ displacements results in a macroscopic polarization and a polar space group ($Pmc2_1$), as shown in Figure \ref{P_by_layer}b. {\it This non-cancellation is the origin of HIF in this class of materials and a route to turn the vast number of perovskite $Pnma$  antiferroelectric materials into functional ferri-electrics}. 

We now formulate a quantitative theory of HIF that relies only on the properties of bulk ABO$_3$ and A$^\prime$BO$_3$, and use it to derive the rule proposed above and displayed in Figure~\ref{tolvs}(d)  that allows for the simultaneous design of a large $P$ and low FE switching barrier.

\subsection{Cation-ordering and the design of high $P$, low switching HIFs.}

Notice above that we implicitly discussed the origin of the spontaneous polarization in this class of materials  in the language of symmetry-lowering distortions of the cubic, 5-atom perovskite structure, that is, A-site cation ordering,  antiferroelectric displacements, and of course the a$^-$a$^-$c$^+$ rotations. A natural reference structure about which to model the polarization in the SLs  is therefore the cubic  perovskite. This has been the essential problem with previous discussions of HIF: all have used the ten atom $P4/mmm$ cation-ordered  perovskite as the reference structure (the ``problem'' will become clear shortly).

\begin{figure}[t]
\centering
\includegraphics[width=9.0cm]{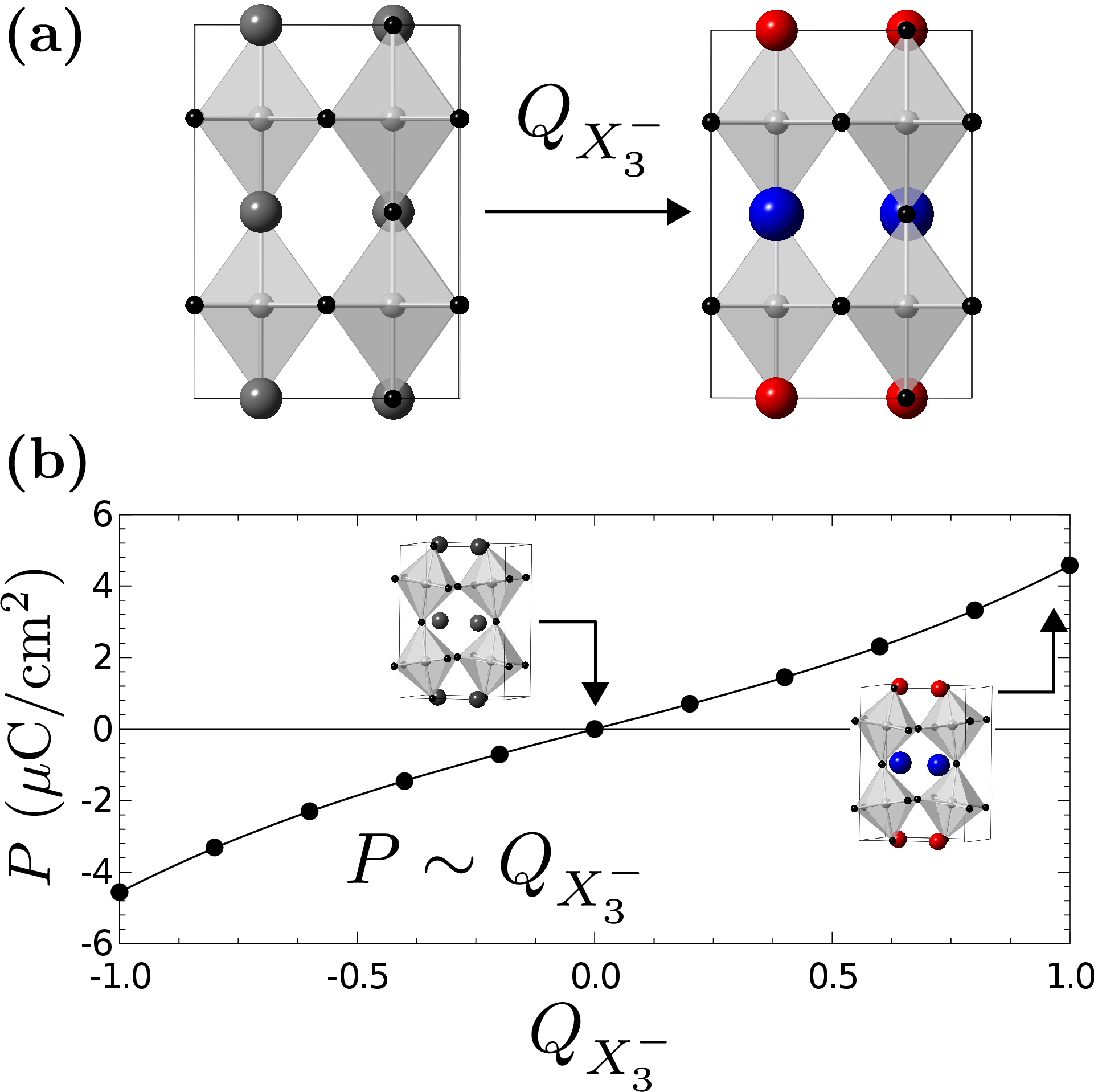}\vspace{-8pt}
\caption{(a) Layered A-site cation ordering mode. (b)  Berry phase polarization as a function of cation ordering calculated within the virtual crystal approximation for  (Sr/Ba)Sn$_2$O$_6$. Note $Q_{X3^-}=0 \Rightarrow$ A-site is fully disordered ($\textit{Pnma}$).}
\label{cation_order}
\end{figure}

Imagine instead that one was to write down an expansion of the total energy in terms of the symmetry-adapted modes that describe the transition from the cubic  $Pm\bar{3}m$  structure to the orthorhombic $Pnma$ structure with the additional stipulation that the A-site is dynamically occupied by two atoms, A and A$^\prime$, with equal probability (hence we are considering the reference structure as a dynamically disordered phase. To be able to describe the layered SL we therefore have to introduce an order parameter to account for the spontaneous ordering of A and A$^\prime$ cations into layers. Group-theoretic analysis shows that this order-disorder transition, shown in Figure~\ref{cation_order}(a), lowers the symmetry of $Pm\bar{3}m$ to $P4/mmm$ through a `composition mode' transforming as  $X_3^-$. 
Further application of group-theoretic techniques shows that in addition to the invariants that describe the $Pm\bar{3}m \longrightarrow Pnma$ transition (all the symmetry allowed couplings of $Q_M$, $Q_R$, and $Q_{X^+_5}$), two nontrivial invariants are introduced into the free energy of $Pm\bar{3}m$: a quadrilinear term,  
\begin{equation}
\mathcal{F}_{MRX_3^-P} = \tilde{\gamma}_1 \tilde{Q}_M \tilde{Q}_R \tilde{Q}_{X_3^-} P,
\label{eq3}
\end{equation}
 (where the tildes refer to modes and coupling constants of the 5-atom, disordered $Pm\bar{3}m$ structure) and an additional trilinear term 
 \begin{equation}
 \mathcal{F}_{X_5^+X_3^-P} = \tilde{\gamma}_2 \tilde{Q}_{X_5^+} \tilde{Q}_{X_3^-} P.
 \label{eq4}
 \end{equation}
We emphasize two points. First, if not for a finite equilibrium value of $\tilde{Q}_{X^-_3}\equiv \langle\tilde{Q}_{X^-_3}\rangle$, these nontrivial couplings between $Pnma$ distortions and the polarization would be symmetry forbidden and therefore hybrid improper ferroelectricity would not possible. We show this explicitly from first-principles calculations of  $P$ in the a$^-$a$^-$c$^+$ structure as a function of cation-ordering, $\langle\tilde{Q}_{X^-_3}\rangle$, within the virtual crystal approximation. This first-principles result, shown in Fig.~\ref{cation_order}(b), shows that as the A-site ordering fully saturates the polarization is maximized, validating our interpretation derived from symmetry arguments. Second, our analysis reveals that there are in fact two contributions to the total polarization: the first 
\begin{equation}
P_{1} \propto \tilde{\gamma}_1\langle\tilde{Q}_{X^-_3}\rangle  \tilde{Q}_M \tilde{Q}_R,
\label{P1}
\end{equation}
originates from the coupling to rotations (Eq.~\ref{eq3}) and is the contribution to $P$ usually thought of when discussing hybrid improper ferroelectricity. The second 
\begin{equation}
P_{2} \propto \tilde{\gamma}_2\langle\tilde{Q}_{X^-_3}\rangle  \tilde{Q}_{X^+_5},
\label{P2}
\end{equation}
originates from a direct coupling to the antipolar displacements, Eq.~\ref{eq4} (a third minor point is to notice that when $X_3^-$ condenses, e.g.,  when one uses the 10-atom ordered system as a reference structure,  the unit cell doubles along [001] and the $\tilde{\gamma}_2$ coupling causes the polar mode  to mix with the antipolar mode resulting in a loss of distinction between ferroelectric and antiferroelectric distortions).

%
Rather than just the direct rotation contribution to the polarization, a surprising consequence of our analysis shows that in general there is a second independent contribution originating from antiferroelectricity, suggesting hybrid improper ferroelectricity could exist without rotations. The identification of a material with rotation-less hybrid improper ferroelectricity would be an exciting new avenue to pursue multifunctional materials as the switching barriers would be expected to be much lower than in proper ferroelectrics.

Antiferroelectricity without  rotations, however, is rare (if it exists at all) in perovskites. Indeed, as we discussed in a previous section, antiferroelectricity in $Pnma$ perovskites is induced by the a$^-$a$^-$c$^+$ rotations.  In such classes of materials our analysis  may therefore be rewritten in the form of an effective trilinear coupling. We  integrate out the antiferroelectric mode 
\begin{equation}
\tilde{Q}_{X^+_5} = -{\tilde{\beta}\over \tilde{A}_{X_5}} \tilde{Q}_M\tilde{Q}_R - {\tilde{\gamma}_2\over \tilde{A}_{X_5}}  \tilde{Q}_{X_3^-} P,
\end{equation}
to lowest order (where $\tilde{A}_{X_5}$ is the force constant of the antiferroelectric mode renormalized by rotations and other distortions), and minimize over $Q_{X^-_3}$  to obtain 
\begin{equation}
\mathcal{F}_{\rm tri}= \gamma \tilde{Q}_M \tilde{Q}_R P
\label{eq5}
\end{equation} 
with $\gamma=     \left( -{\tilde{\gamma}_2  } { \tilde{\beta}\over \tilde{A}_{X_5}}+  \tilde{\gamma}_1  \right)  \langle \tilde{Q}_{X^-_3}\rangle$ (see Appendix  for details), which leads to
\begin{equation}
 \Rightarrow P=\gamma {\langle\tilde{Q}_M\rangle \langle\tilde{Q}_R\rangle \over \tilde{A}_P}.
\label{eq5b}
\end{equation} 

This analysis makes clear that the trilinear coupling coefficient, $\gamma$, is composed of two factors: the magnitude of the cation ordering, $\langle\tilde{Q}_{X^-_3}\rangle$ and a term that goes like $-1/(\tilde{A}_{X_5} - 1)$. Notice that the latter term differs from $1$ due to the direct coupling of the antipolar and polar distortions. Therefore, if $\tilde{A}_{X_5}$  has a dependence on tolerance factor, then even in  systems  where antiferroelectricity by itself is stable (such as the majority of the $Pnma$ perovskites we have been discussing), the conjecture that the polarization is simply proportional to  the amplitude of the rotations is incorrect.  Although interesting, the consequences of this are not profound as any physically reasonable force constant would be a smooth function of the tolerance factor and therefore cannot be the origin of the discrepancy displayed in Figure~\ref{tolvs}(d).

What is interesting, is the only remaining unknown -- the cation ordering $\langle\tilde{Q}_{X^-_3}\rangle$. One normally thinks of this as a number between 0 and 1,  reflecting the crystallographic difference in occupancy of a site. Here this occupancy difference is irrelevant in the sense that in the ordered state of the $Pnma$ SLs, the important detail (in relation to $P$) is the difference in the `susceptibility' of the A and A$^\prime$ cations to displace from their ideal $Pm\bar{3}m$ positions~\cite{discrep}. This can be made clear by considering the following example. Imagine two different A/A$^\prime$  cation-ordered materials that are both 100$\%$ ordered,  but in one case the tendency of the A and A$^\prime$ cations  to off-center is similar, whereas in the second system the A and A$^\prime$  tendency to off-center varies drastically. Even though cation ordering is 100$\%$ in each of these systems,  the consequences of this ordering  should  differ considering the ferrielectric mechanism  described. As far as the polarization is concerned it is irrelevant that A and A$^\prime$ may be different atoms only to have their tendency to off-center  be similar as this would lead to a near complete cancelation of the layered dipoles (just as if they were the same atom). {\it In this case, it is therefore fruitful to think of the difference in the ability  to displace  between two A-sites as an absolute magnitude of the cation ordering.}

We propose that the absolute magnitude of the cation ordering is reflected in the  difference in  magnitude of the antiferroelectric mode, $Q_{X^+_5}$,  observed in the ABO$_3$ perovskite from that observed in A$^\prime$BO$_3$,
\begin{equation}
\langle\tilde{Q}_{X^-_3}\rangle \propto \left(\langle Q_{X^+_5}^{ABO_3}\rangle - \langle Q_{X^+_5}^{A'BO_3}\rangle \right) \equiv \Delta_{Q_{X5^+}}.
\end{equation}
Since  $Q_{X^+_5}$ is linear in $1-\tau$ (see appendix), 
\begin{equation}
\Delta_{Q_{X5^+}} \propto \Delta{\tau} 
\end{equation}
becomes a simple crystal chemistry rule to understand how the cation ordering contribution to $P$ is related to the difference in tolerance factor, $\Delta\tau$, between ABO$_3$ and A$'$BO$_3$.~\cite{rodriguez96,sinclair99}

\begin{figure*}[t]
\centering
\includegraphics[width=16.0cm]{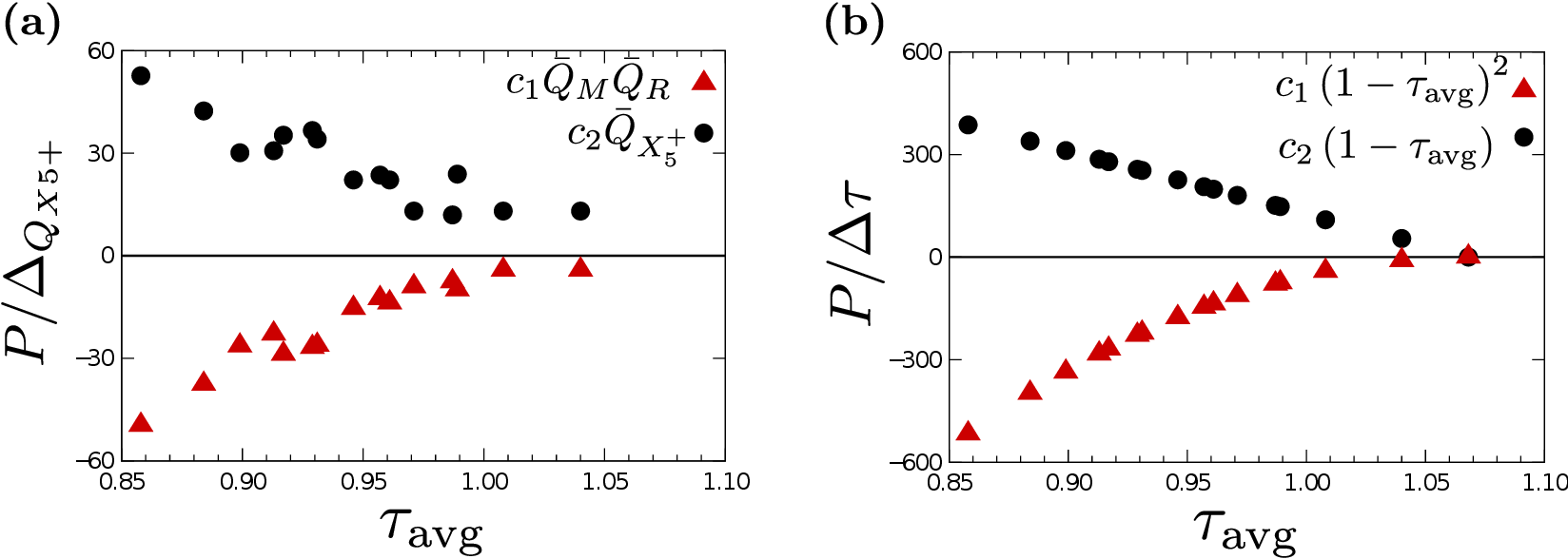}\vspace{-0pt}\\
\caption{(a) Polarization contributions from fitting to first principles mode amplitudes scaled by the difference in magnitude of the antiferroelectric mode, (b) polarization contributions from tolerance factor model scaled by the difference in tolerance factor. }
\label{model}
\end{figure*}

%
Assuming that the $\tilde{Q}_n$'s in the disordered  perovskite can be approximated by the average of the respective $Q_n$'s in each constituent, \textit{i.e.}, $\langle\tilde{Q}_n\rangle\approx\bar{Q}_n$, 
\begin{equation}
{\gamma}\propto  ( -{\tilde{\gamma}_2  } { \tilde{\beta}\over 1-\tau_{\rm avg}}+  \tilde{\gamma}_1  )      \Delta{\tau}
\end{equation}
 and  $\langle\tilde{Q}_M\rangle \langle\tilde{Q}_R\rangle\propto (1-\tau_{\rm avg})^2$  we therefore have 
\begin{eqnarray}
P&=&   -{1\over \tilde{A}_P}{\gamma} \langle\tilde{Q}_M\rangle \langle\tilde{Q}_R\rangle \nonumber \\
  & \propto &( -{\tilde{\gamma}_2  } { \tilde{\beta}\over 1-\tau_{\rm avg}}+  \tilde{\gamma}_1  )   \Delta{\tau} (1-\tau_{\rm avg})^2
\label{eq6}
\end{eqnarray}
and therefore
\begin{equation}
P=  \Delta{\tau} \left(c_1 (1-\tau_{\rm avg})^2 + c_2 (1-\tau_{\rm avg}) \right)
\label{Pmodel}
\end{equation}
which  shows a nearly perfect fit to the first-principles calculations across all three families of compounds as discussed previously and shown in Figure~\ref{tolvs}(c). In Figure~\ref{model} we show each contribution to the polarization separately across our material suite. Each contribution, computed using only the first-principles structural distortion amplitudes, is a smooth monotonic function of tolerance factor when scaled by the difference in magnitude of the antiferroelectric mode. Furthermore, the rotation-driven term from Equation~\ref{P1} has an oppositely oriented polarization than the antiferroelectric-driven term from Equation~\ref{P2}. Notice that  the rotation-driven polarization term dominates for small tolerance factor materials. Quite surprisingly, however, it is the antiferroelectric contribution that dominates for materials with $\tau_\textrm{avg}$ close to 1 where the  switching barrier is small.

This quantitative model results in a simple, chemically intuitive design rule of thumb.
\begin{itemize}
\item []{\it To keep the switching barrier low and the polarization high: two bulk ABO$_3$ and A$^\prime$BO$_3$ perovskites should be selected such that the average tolerance factor, $\tau_\textrm{avg}$, is maximized and the difference in their tolerance factors, $\Delta\tau$, is also large,}
\end{itemize}
and for most reasonable tolerance factors the model is equivalent to
\begin{equation}
P\sim  \Delta{\tau}   (1-\tau_{\rm avg}).
\end{equation}
%

\subsection{Extension to Perovskite-derived cation ordered Ruddlesden-Popper materials}

\begin{figure}[b]
\includegraphics[width=8cm]{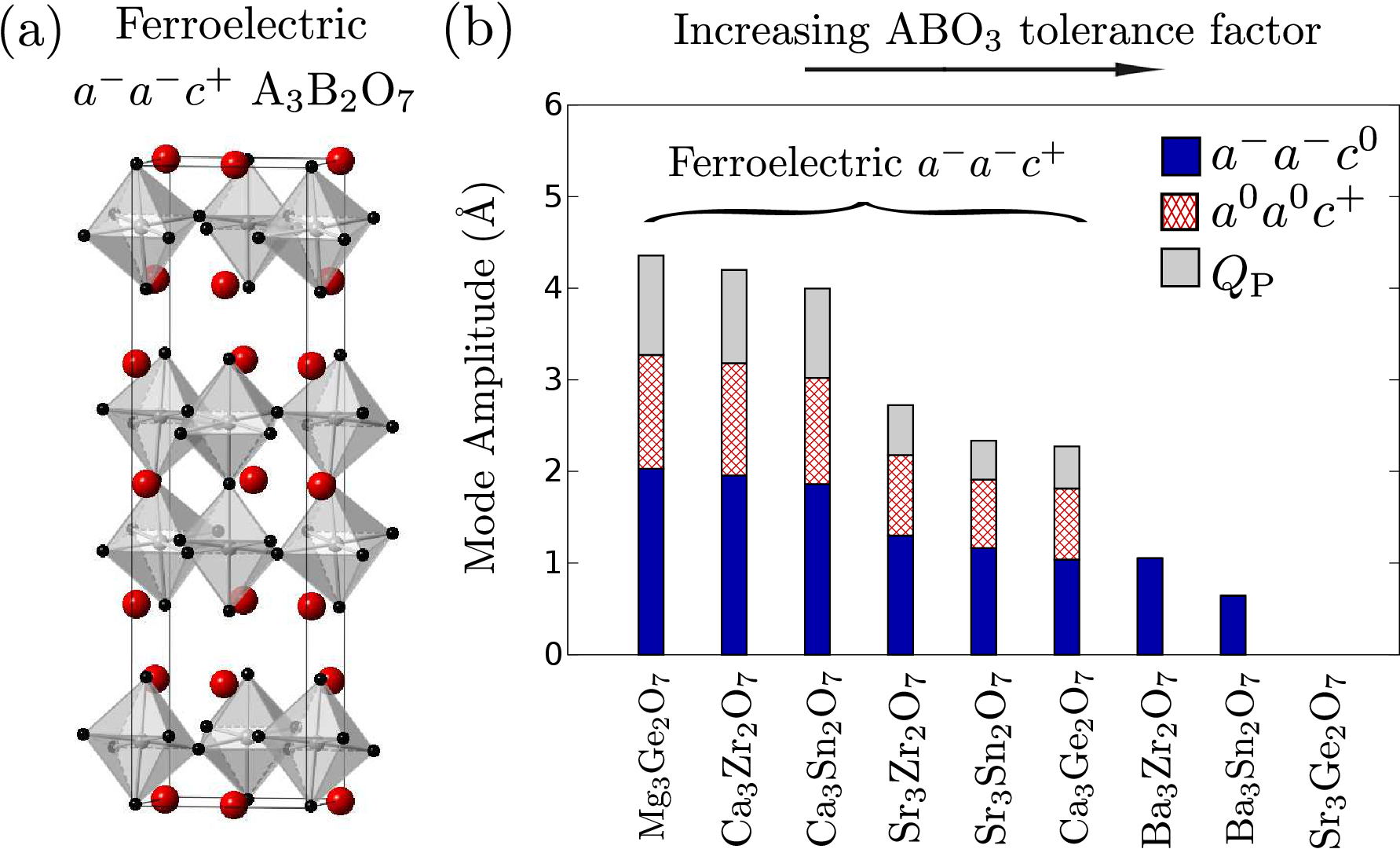}
\caption{(a) The ferroelectric $a^-a^-c^+$ structure of the A$_3$B$_2$O$_7$ Ruddlesden-Popper phase. (b) First principles amplitudes of the two octahedral rotations and the induced polar mode for a suite of A$_3$B$_2$O$_7$ materials, arranged by increasing tolerance factor of the ABO$_3$ parent.}
\label{RP_1}
\end{figure}

The chemical intuition learned in the previous sections is applicable to more than just perovskite heterostructures having a  3D network of corner-connected octahedra such as the 1-1 SL.  For example, it has been known for some time that  certain $n=2$ Ruddlesden-Popper (RP) compounds -- layered perovskites with  \emph{disconnected} octahedra along [001] --    display rotations of their octahedra  and are polar~\cite{aleksandrov95}, as shown in Figure \ref{RP_1}(a). Recently, one such system, Ca$_3$Mn$_2$O$_7$~\cite{benedek11, harris11}, has been shown to display a strong coupling between the rotation-induced ferroelectricity and magnetism. 
An analysis by Aleksandrov and colleagues has suggested that the $a^-a^-c^+$ octahedral rotation pattern is responsible for the polar state. Is this really true?

Using our suite of ABO$_3$ perovskites we studied nine A$_3$B$_2$O$_7$ systems spanning a wide range of ABO$_3$ tolerance factors. In Figure \ref{RP_1}(b) we plot the amplitudes of the $a^-a^-c^0$ and $a^0a^0c^+$ distortion patterns. For the six A$_3$B$_2$O$_7$ derived from $Pnma$ perovskites, we find that the RP structure displays a similar rotation pattern and is polar. This indeed suggests that any $Pnma$ perovskite  would be polar if it can be synthesized in the $n=2$, A$_3$B$_2$O$_7$ RP structure.  

\begin{figure}[t]
\includegraphics[width=7.0cm]{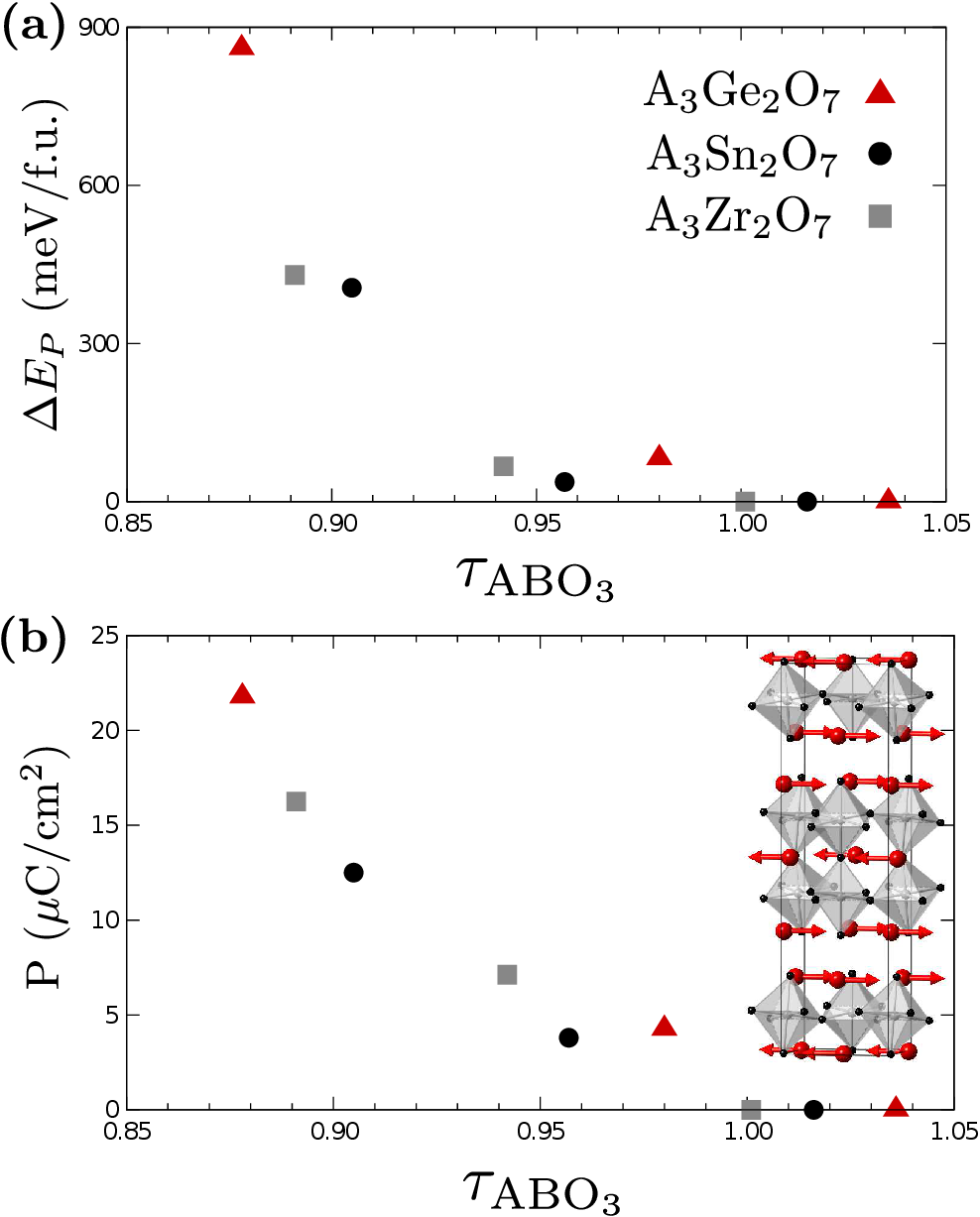}
\caption{(a) Stabilization energy, and (b) polarization of the $n=2$ Ruddlesden-Popper materials arranged by the tolerance factor of the ABO$_3$ end member. Inset: the A-cation displacement pattern in the ferroelectric $a^-a^-c^+$ structure.}
\label{RP_2}
\end{figure}

In order to realize a functional ferroelectric, however, the polarization should be large with a low switching barrier.
Based on what we learned in the previous section we would expect that as the tolerance factor of the ABO$_3$ perovskite making up the RP material decreases, the octahedral rotations, and consequently the polarization and the switching barrier, would get larger. As shown in Figures~\ref{RP_1}(b), Figure~\ref{RP_2}(a), and Figure~\ref{RP_2}(b), this is exactly the case. Perhaps the large barrier to switch the polarization in Ca$_3$Mn$_2$O$_7$ or Ca$_3$Ti$_2$O$_7$ (two known polar RPs)  is the reason why ferroelectricity has yet to be shown experimentally. (Note, Sr$_3$Sn$_2$O$_7$, which has been synthesized previously,~\cite{green00} is an ideal system to explore for ferroelectricity).

Is the origin of the polar state in these $Pnma$ RPs similar to what we have been discussing? First note that unlike the perovskite superlattices, the polarization is a smooth and monotonic function of tolerance factor. 
This  behavior is  in fact consistent with the mechanism we outlined in the previous sections and can be easily understood by examining the RP structure. First, in even layered RPs, such as the $n=2$, the interface between the rock salt layer and the perovskite layer breaks the local inversion symmetry of the individual octahedra, therefore cation-ordering in not required as in the perovskite superlattices.  Additionally, as shown in the inset to Figure~\ref{RP_2}(b), the $a^-a^-c^+$ octahedral rotation pattern  still  induces antipolar A-cation displacements in the RP. Given that the AO layers at the interface and the layer between the perovskites are different by symmetry, a net polarization will arise -- {\it hybrid improper ferroelectricity without cation ordering}.
Unlike the superlattice example however, the noncancellation of the A-cation displacements from this effect is very small, yet there is still a substantial polarization because of the additional AO layer present in the unit cell of the RP. The polarization  varies smoothly because both of these contributions depend only on one type of A-cation antiferroelectric displacement and therefore only on one ABO$_3$ tolerance factor.

\begin{figure}[t]
\includegraphics[width=8cm]{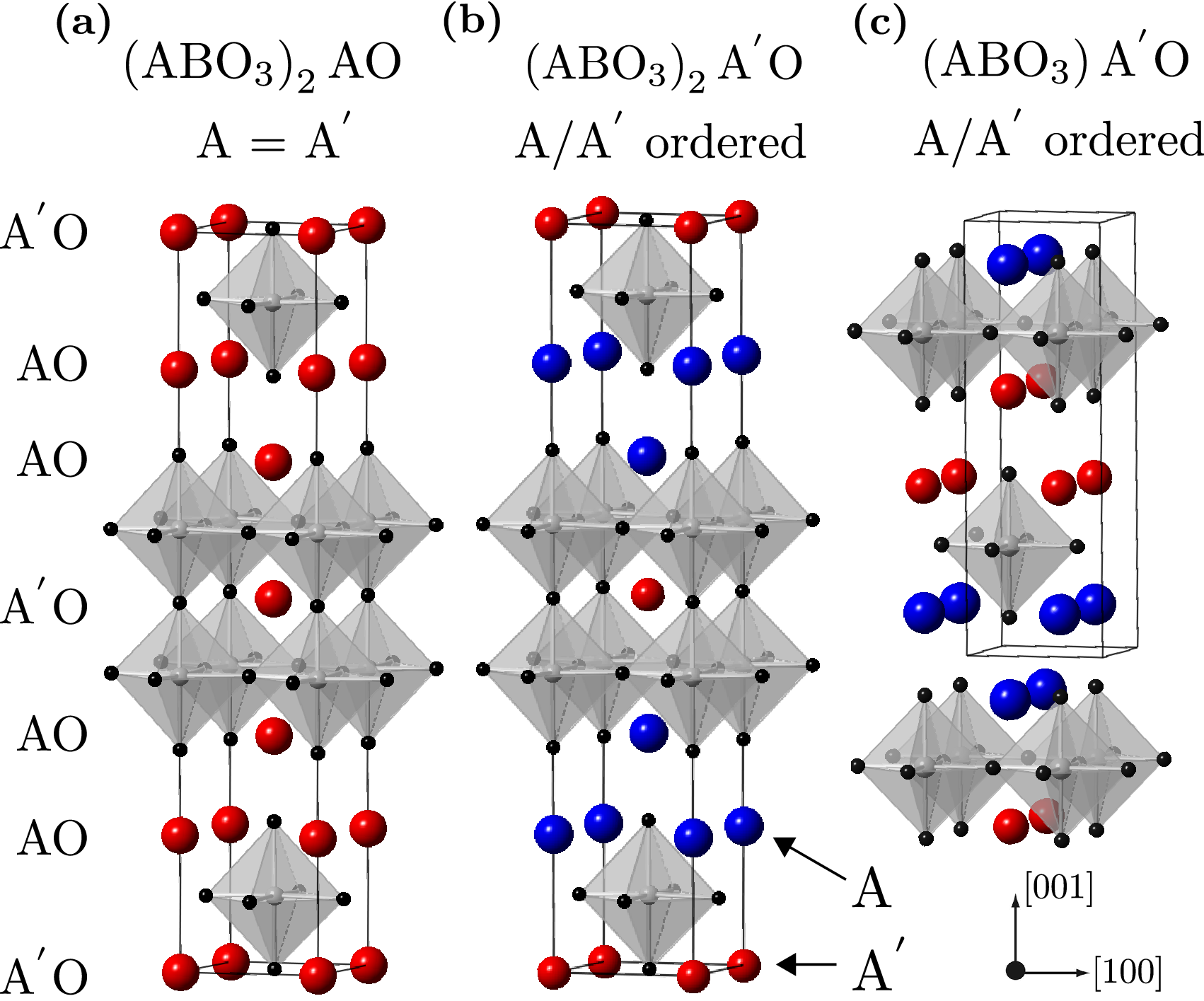}
\caption{(a) The $n=2$ Ruddlesden-Popper structure (without octahedral rotations) with a single A-cation. (b) A/A' ordered $n=2$ phase, (c) A/A' ordered $n=1$ phase. All three of these layered perovskites are polar in the $a^-a^-c^+$ ground state.}
\label{RP_3}
\end{figure}

Having established a common origin, can we use the chemical mismatch strategy, $\Delta \tau$, learned in the previous section, to reduce the switching barrier (by increasing $\tau_{\rm avg}$)  while increasing the polarization in the A$_3$B$_2$O$_7$ structure? The design rules tell us to create A/A' cation ordered RP structures, as shown in Figure \ref{RP_3}. Tables \ref{superRP} and \ref{superRP_Zr} show the first-principles polarization and switching barrier in the $\left(\text{ASnO}_3\right)_2$A'O and $\left(\text{AZrO}_3\right)_2$A'O materials, respectively. Taking $\left(\text{CaSnO}_3\right)_2$CaO as an example, and reading down the first column of Table \ref{superRP}, we find
\begin{itemize}
\item[]   $\left(\text{CaSnO}_3\right)_2$CaO, P=12 $\mu$C/cm$^2$, $\Delta$E=406 meV  \\
 \textcolor{white}{qqqqqqqqqq}$\searrow$\\ 
 \textcolor{white}{q} $\left(\text{CaSnO}_3\right)_2$SrO, P=14 $\mu$C/cm$^2$, $\Delta$E=234 meV \\
  \textcolor{white}{qqqqqqqqqqqqq}$\searrow$\\ 
\textcolor{white}{qqq} $\left(\text{CaSnO}_3\right)_2$BaO, P=17 $\mu$C/cm$^2$, $\Delta$E=122 meV  \\
\end{itemize}
which is remarkably consistent with our superlattice design rules.
While substitution of the A' cation (in between the perovskite blocks) in $\left(\text{CaSnO}_3\right)_2$CaO reduces the switching barrier and increases the polarization, Table~\ref{superRP} shows that substitution of the A cation (in the rock salt layer) reduces the barrier but also \textit{decreases} the polarization. For example, reading across the first row we find
\begin{itemize}
\item[]   $\left(\text{CaSnO}_3\right)_2$CaO, P=12$\mu$C/cm$^2$, $\Delta$E=406 meV  \\
 \textcolor{white}{q}$\searrow$\\ 
 \textcolor{white}{q} $\left(\text{SrSnO}_3\right)_2$CaO, P=3.4$\mu$C/cm$^2$, $\Delta$E=182 meV \\
  \textcolor{white}{qqq}$\searrow$\\ 
\textcolor{white}{qqq} $\left(\text{BaSnO}_3\right)_2$CaO, P=3.5$\mu$C/cm$^2$, $\Delta$E=145 meV  \\
\end{itemize}
Does this contradict our chemical mismatch design strategy? 

\begin{figure}[t]
\includegraphics[width=8.8cm]{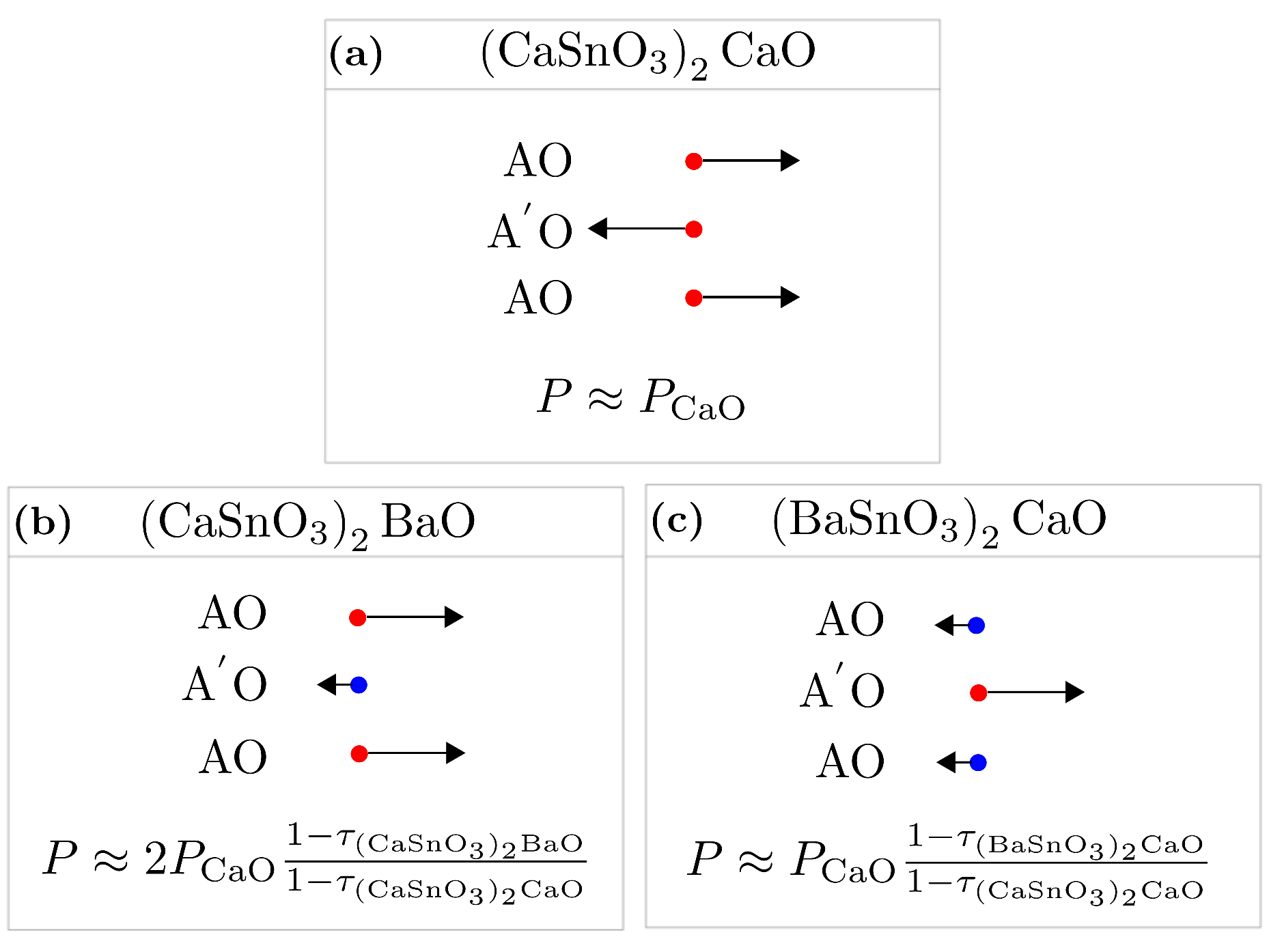}
\caption{Simple model for the polarization showing the AO and A$^{'}$O layer contributions. (a) A=A$^{'}$, (b) A$^{'}$ is replaced by a larger cation (for instance Ba replaces Ca)  (c) A is replaced by a larger cation.}
\label{RP_4}
\end{figure}

Let us first look at $\left(\text{CaSnO}_3\right)_2$CaO where A=A$^{'}$. As Figure~\ref{RP_4}(a) shows, because of the weak symmetry breaking of the perovskite/rocksalt interface, the polarization originating from one of the AO layers nearly cancels that coming from the A$^{'}$O layer and therefore in this material the polarization is  approximately that originating from one CaO layer. 
Now with the substitution of the A$^{'}$ site with a different cation, as shown in Figure~\ref{RP_4}(b) the AO layer and the A$^{'}$O layer no longer cancel. In the case where  the A$^{'}$ cation radius is much larger, for example A$^{'}$=Ba, the polarization in this  A$^{'}$O layer is nearly zero. According to the model we derived in the previous sections, this should lead to a  polarization that is approximately twice the contribution from one CaO layer multiplied by the ratio of the average ABO$_3$ tolerance factors, i.e., $P\approx$ 15$\mu$C/cm$^2$, which compares amazingly well to the first-principles result of 17$\mu$C/cm$^2$.
 Going across the row, the tolerance factor increases much faster than the mismatch increase and the polarization decreases, e.g., the model  predicts $P\approx$2.3$\mu$C/cm$^2$ (first-principles value equals 3.5 $\mu$C/cm$^2$) in $\left(\text{BaSnO}_3\right)_2$CaO  because the polarization from one CaO layer is reduced by the substantial increase in the average tolerance factor as shown in Figure~\ref{RP_4}(c).
%

The same general design rules can be applied to the $n=1$ member of the Ruddlesden-Popper series. Although the $a^-a^-c^+$ rotation pattern does not induce ferroelectricity in A$_2$BO$_4$ as there are an \emph{even} number of A cations per perovskite block, the $a^-a^-c^+$ ground state of the A/A' cation ordered $\left(\text{ABO}_3\right)$AO is polar \cite{benedek12}, for example: 
\begin{itemize}
\item[]   $\left(\text{CaSnO}_3\right)$SrO, P = 5.7 $\mu$C/cm$^2$, $\Delta$E = 235 meV  \\
 \textcolor{white}{qqqqqqqqqq}$\searrow$\\ 
 \textcolor{white}{qq} $\left(\text{CaSnO}_3\right)$BaO, P = 11 $\mu$C/cm$^2$, $\Delta$E = 159 meV \\
\end{itemize}
which is consistent with our general design rules.
%

\begin{table}[b]
\caption{Ground state total polarization and switching energy barrier of the six possible $\left(\text{ASnO}_3\right)_2\text{A'O}$ superlattices. ``--''  indicates compounds are stable in the paraelectric structure without any octahedral rotation about the [001] axis.}
\centering
\renewcommand{\tabcolsep}{2mm} 
\renewcommand{\arraystretch}{1.5} 
\begin{ruledtabular}
\begin{tabular}{c c c c }
A'O & $\left(\text{CaSnO}_3\right)_2$A'O & $\left(\text{SrSnO}_3\right)_2$A'O & $\left(\text{BaSnO}_3\right)_2$A'O \\
\hline \hline
\multirow{2}{*}{CaO} & {\color{black}12 $\mu$C/cm$^2$}  & {\color{black}3.4 $\mu$C/cm$^2$}  & {\color{black}3.5 $\mu$C/cm$^2$}  \\
 & {\color{black}406 meV} & {\color{black}182 meV} & {\color{black}145 meV} \\
 \hline
\multirow{2}{*}{SrO} & {\color{black}14 $\mu$C/cm$^2$}  & {\color{black}3.9 $\mu$C/cm$^2$}  & {\color{black}1.8 $\mu$C/cm$^2$}  \\
 & {\color{black}234 meV} & {\color{black}37 meV} & {\color{black}10 meV} \\
 \hline
\multirow{2}{*}{BaO} & {\color{black}17 $\mu$C/cm$^2$}  & \multirow{2}{*}{-}  & \multirow{2}{*}{-}  \\
& {\color{black}122 meV} &  &  \\
\end{tabular}
\end{ruledtabular}
\label{superRP}
\end{table}

%
\section{conclusion}
%

We have shown how to take the most common perovskite, the $Pnma$ perovskite, and turn it into a functional ferroelectric.   We have presented the microscopic theory of this new type of ferroelectricity, based on the interplay of steric octahedral rotations, antiferroelectricity, and cation ordering, through both group theoretical methods and systematic first-principles calculations on more than 16 different ordered ABO$_3$/A$^{'}$BO$_3$ superlattices and 20 different $\left(\text{ABO}_3\right)_2$A$^{'}$O and $\left(\text{ABO}_3\right)_1$A$^{'}$O Ruddlesden-Popper (RP) compounds.
Using results  only on superlattices, we have derived a design strategy expressed only in terms of crystal-chemistry descriptors of the ABO$_3$ constituents. We have shown that these straightforward design concepts also capture the basic physics of the RP systems as well and are likely general, universal concepts.
There are no chemical restrictions on the type of A and/or B cations that can be used, making this ferroelectric mechanism compatible with the types of correlated electron phenomena, for example magnetism, usually associated strongly with the $Pnma$ perovskites. 
This materials design framework for achieving large electrical polarizations and low switching barriers will be a crucial guide to the successful integration of these new multifunctional ferroelectric materials into next generation devices.

\begin{table}[t]
\caption{Ground state total polarization and switching energy barrier of the six possible $\left(\text{AZrO}_3\right)_2\text{A'O}$ superlattices.}
\renewcommand{\tabcolsep}{2mm} 
\renewcommand{\arraystretch}{1.5} 
\begin{ruledtabular}
\begin{tabular}{c c c c }
A'O & $\left(\text{CaZrO}_3\right)_2$A'O & $\left(\text{SrZrO}_3\right)_2$A'O & $\left(\text{BaZrO}_3\right)_2$A'O \\
\hline \hline

\multirow{2}{*}{CaO} & {\color{black}16 $\mu$C/cm$^2$}  & {\color{black} 6.0 $\mu$C/cm$^2$}  & {\color{black} 4.4 $\mu$C/cm$^2$}  \\
 & 430 meV & 192 meV & 129 meV \\
 \hline

\multirow{2}{*}{SrO} & {\color{black} 17 $\mu$C/cm$^2$}  & {\color{black}7.2 $\mu$C/cm$^2$} & {\color{black} 3.4 $\mu$C/cm$^2$} \\
 & 293 meV & 67 meV & 19 meV \\
 \hline

\multirow{2}{*}{BaO} & {\color{black} 24 $\mu$C/cm$^2$}  & {\color{black} 11 $\mu$C/cm$^2$} & \multirow{2}{*}{-} \\
 & 193 meV & 10 meV & \\

\end{tabular}
\end{ruledtabular}
\label{superRP_Zr}
\end{table}

\section{method}
First-principles calculations are performed within density functional theory as implemented in {\sc Quantum Espresso}~\cite{QE-2009} using ultrasoft pseudopotentials and the PBEsol functional~\cite{perdew08}, which provides an improved description of structure~\cite{perdew08} over LDA or PBE~\cite{roman08}. We used a plane wave energy cutoff of 650 eV. We used a \textbf{k}-point grid equivalent to an 8$\times$8$\times$8 grid of the cubic perovskite. The total polarization was calculated using the Berry phase method.

\section{acknowledgements}
We acknowledge useful discussions with Pat Woodward. A.T.M$.$ was supported by NSERC of Canada and by the NSF (No. DMR-1056441). J.M.R$.$ was supported by ARO, under Contract Number W911NF-12-1-0133. N.A.B$.$ and  C.J.F$.$  were supported by DOE-BES under Award Number DE-SCOO02334. \\

\section{Appendix: Landau Theory}

The choice of high symmetry reference structure is a key step in modelling materials in which multiple distortion modes couple in nontrivial ways. Prior work on hybrid improper ferroelectrics used the A/A' cation ordered \textit{P4/mmm} structure, rather than the cation disordered cubic \textit{Pm$\bar{3}$m}, as the high symmetry reference structure. In this appendix we provide the Landau theory for each of these reference structures.

\subsection{Polarization Starting from the Cation Ordered \textit{P4/mmm} Structure}

\begin{figure}[b]
\centering
\includegraphics[height=5cm]{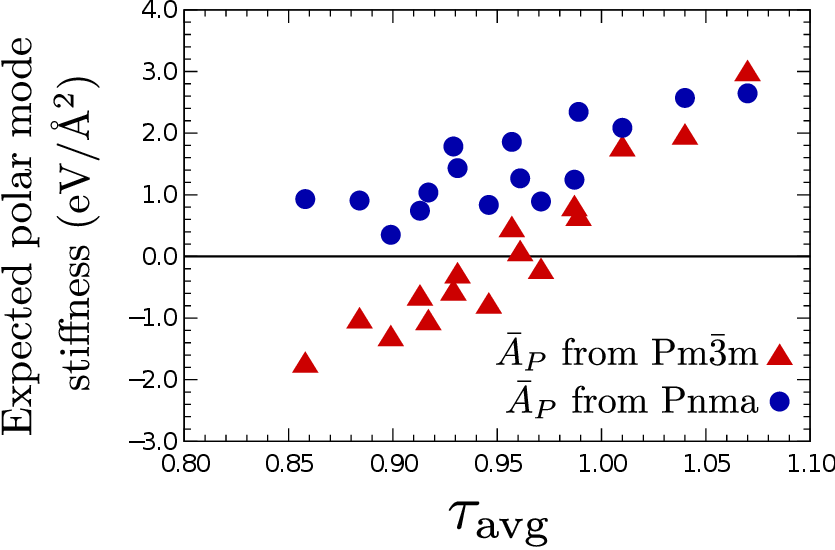}
\caption{The average polar mode stiffness (force constant) of ABO$_3$ and A'BO$_3$ in the high symmetry Pm$\bar{3}$m structure and renormalized in the Pnma $a^-a^-c^+$ structure. }
\label{supp_polarmode}
\end{figure}

In the 10 atom, cation ordered \textit{P4/mmm} structure (see Figure \ref{cation_order}a) there are three key distortion modes that we will treat as order parameters: $Q_{R1}$ which describes the strength of $a^-a^-c^0$ octahedral rotations, $Q_{R2}$ which describes the strength of $a^0a^0c^+$ octahedral rotations, and the polarization. The free energy around the \textit{P4/mmm} structure is
\begin{equation}
\begin{split}
\mathcal{F} &= \frac{1}{2}\sum_{i}A_{i} Q_{i}^2 + \frac{1}{4}\sum_{i}B_{i} Q_{i}^4\\ &+ \frac{1}{2}\sum_{ij}B_{ij} Q_{i}^2 Q_{j}^2 + \gamma Q_{R1} Q_{R2} P \\
\end{split}
\label{F-P4mmm}
\end{equation}
where the summation index denotes the three order parameters $Q_{R1}$, $Q_{R2}$, and $P$. All of the ferroelectric materials we consider have $A_{R1},A_{R2} < 0$. In other words, the \textit{P4/mmm} structure is unstable with respect to the $a^-a^-c^+$ rotation pattern.  We set $Q_{R1} = \left<Q_{R1}\right>$ and $Q_{R2} = \left<Q_{R2}\right>$ to obtain the effective free energy in the polar $a^-a^-c^+$ structure.
\begin{equation}
\begin{split}
\mathcal{F} &= \frac{1}{2}\tilde{A}_{P} P^2 + \gamma \left<Q_{R1}\right> \left<Q_{R2}\right> P \\
\end{split}
\label{p4mmm}
\end{equation}
where $\tilde{A}_P$ is the polar mode stiffness after renormalization from the biquadratic couplings with $\left<Q_{R1}\right>$ and $\left<Q_{R2}\right>$ and any other modes which are symmetry-allowed in the polar structure. In Figure \ref{supp_polarmode} we show how the renormalized polar mode is stable, $\tilde{A}_P > 0$, even when the high symmetry structure has a polar instability. From Equation \ref{p4mmm} we therefore conclude the polarization arises from the trilinear coupling to the octahedral rotations. Minimizing over P leads to the lowest order spontaneous polarization
\begin{equation}
P = \gamma \frac{\left<Q_{R1}\right>\left<Q_{R2}\right>}{\tilde{A}_P} .
\label{ps}
\end{equation}

In the main text Figure 2 we show that $\left<Q_{R1}\right>\left<Q_{R2}\right> \propto 1 - \tau_{\rm avg}$. Figure \ref{supp_polarmode} shows that the renormalized polar mode force constant $\tilde{A}_P$ is roughly independent of tolerance factor, and therefore from this analysis we would expect $P \propto 1 - \tau_{\rm avg}$. As discussed in the main text this tolerance factor dependence is not observed in our first principles calculations. This apparent problem is resolved by using the A/A' cation disordered \textit{Pm$\bar{3}$m} reference structure.

\subsection{Polarization Starting from the Cation Disordered \textit{Pm$\bar{3}$m} Structure}

 Instead of the 10 atom \textit{P4/mmm} structure we will consider the Landau free energy expanded around the 5 atom cation disordered \textit{Pm$\bar{3}$m} structure. The \textit{P4/mmm} space group is then a subgroup of \textit{Pm$\bar{3}$m}, connected by the irrep $X_3^-$ which we physically attribute to cation ordering of the A-site (see Figure 5). We will now demonstrate how A-site ordering leads to an effective trilinear coupling between a$^-$a$^-$c$^+$ rotations and the polarization. The spontaneous polarization will be given by Equation \ref{ps} except that the \textit{P4/mmm} trilinear coupling, $\gamma$, will be replaced by an effective coupling which will depend on the magnitude of the A/A' antiferroelectric displacements. 

In the 5-atom, cation disordered \textit{Pm$\bar{3}$m} structure there are five relevant mode amplitudes that we take to be the order parameters in our Landau theory: $\tilde{Q}_{M}$, $\tilde{Q}_{R}$, $\tilde{Q}_{X_5^+}$, $\tilde{Q}_{X_3^-}$, and P (see main text for details). To fourth order the free energy around the \textit{Pm$\bar{3}$m} structure is
\begin{equation}
\begin{split}
\mathcal{F} &= \frac{1}{2}\sum_{i}A_{i}\tilde{Q}_{i}^2 + \frac{1}{4}\sum_{i}B_{i}\tilde{Q}_{i}^4 \\ &+ \frac{1}{2}\sum_{ij}B_{ij}\tilde{Q}_{i}^2\tilde{Q}_{j}^2 + \tilde{\beta} \tilde{Q}_{M}\tilde{Q}_{R}\tilde{Q}_{X_5^+} \\ &+ \tilde{\gamma}_2 \tilde{Q}_{X_5^+}\tilde{Q}_{X_3^-}P + \tilde{\gamma}_1\tilde{Q}_{M}\tilde{Q}_{R}\tilde{Q}_{X_3^-}P \\
\end{split}
\label{fullF}
\end{equation}

where the summation index denotes the five order parameters specified above.

We first integrate out the antipolar mode $\tilde{Q}_{X_5^+}$ to obtain an effective free energy in the remaining order parameters. We assume the rotations sterically induce this antipolar motion and thus $A_{X_5^+} > 0$ (which is true for $\tau \gtrsim 0.91$, see Figure \ref{Fig_X5}). We obtain

\begin{equation}
\tilde{Q}_{X_5^+} = -\frac{\tilde{\beta}}{A_{X_5^+}^{'}}\tilde{Q}_{M}\tilde{Q}_{R} - \frac{\tilde{\gamma}_2}{A_{X_5^+}^{'}}\tilde{Q}_{X_3^-}P
\end{equation}

where $A_{X_5^+}^{'}$ is the mode stiffness after renormalization from the biquadratic couplings of each other mode to $\tilde{Q}_{X_5^+}$ and to lowest order $A_{X_5^+}^{'} = A_{X_5^+}$. Finally, in the cation ordered 10 atom cell (see the inset of Figure 3(c) in the main text) we set $\tilde{Q}_{X_3^-} = \left<\tilde{Q}_{X_3^-}\right>$ to obtain the effective free energy for the three hybrid improper order parameters $\tilde{Q}_{M}$, $\tilde{Q}_{R}$ and P. By substitution of $\tilde{Q}_{X_5^+}$ and $\tilde{Q}_{X_3^-}$ in the free energy of Equation S1, we obtain
\begin{equation}
\begin{split}
\mathcal{F_{\rm Eff}} &= \frac{1}{2}\sum_{i}\tilde{A}_{i}\tilde{Q}_{i}^2 + \frac{1}{2}\sum_{ij}\tilde{B}_{i,j}\tilde{Q}_{i}^2\tilde{Q}_{j}^2 \\
&+ \left(-\tilde{\gamma}_{2}\frac{\tilde{\beta}}{A_{X_5^+}^{'}} + \tilde{\gamma}_{1}\right)\left<\tilde{Q}_{X_3^-}\right>\tilde{Q}_{M}\tilde{Q}_{R}P 
\end{split}
\end{equation}
where the summation index now denotes the two rotation modes and the polarization, and the new effective coefficients $\tilde{A}_{M}$, $\tilde{A}_{R}$, $\tilde{A}_{P}$ and $\tilde{B}_{M,R}$ are modified from the values in Equation S1 through the coupling terms. The effective trilinear coupling constant between $\tilde{Q}_{M}$, $\tilde{Q}_{R}$ and P is

\begin{equation}
\gamma = \left(-\tilde{\gamma}_{2}\frac{\tilde{\beta}}{A_{X_5^+}^{'}} + \tilde{\gamma}_{1}\right)\left<\tilde{Q}_{X_3^-}\right> .
\end{equation}

\subsection{Polar-Antipolar Mixing}

In this section we formally demonstrate how the cation ordering of the A-site mixes the polar and antipolar modes of the constituent ABO$_3$ perovskite. The cation ordering along the [001] direction (equivalent to a ABO$_3$/A$^\prime$BO$_3$ superlattice) is accounted for in our Landau theory by the `composition' order parameter $\tilde{Q}_{X_3^-}$ (see main text for details). In the 10 atom cation ordered cell, the free energy of the disordered perovskite in Equation \ref{fullF} is modified by $\tilde{Q}_{X_3^-} = \left<\tilde{Q}_{X_3^-}\right>$, which leads to an effective free energy of the remaining order parameters,
\begin{equation}
\begin{split}
\mathcal{F}_{\rm Eff} &= \frac{1}{2}\sum_{i}A_{i}\tilde{Q}_{i}^2 + \frac{1}{4}\sum_{i}B_{i}\tilde{Q}_{i}^4 \\
&+ \frac{1}{2}\sum_{ij}B_{ij}\tilde{Q}_{i}^2\tilde{Q}_{j}^2 + \tilde{\beta} \tilde{Q}_{M}\tilde{Q}_{R}\tilde{Q}_{X_5^+} \\
&+ \tilde{\gamma}_2\left<\tilde{Q}_{X_3^-}\right>\tilde{Q}_{X_5^+}P + \tilde{\gamma}_1\left<\tilde{Q}_{X_3^-}\right>\tilde{Q}_{M}\tilde{Q}_{R}P .
\end{split}
\label{bilinear}
\end{equation}
The bilinear coupling of Equation \ref{bilinear} leads to a mixed normal mode in the antipolar and polar distortions, $\tilde{Q}_{X_5^+}$ and P (see Figure 3 in the main text). We can write the mixed mode as

\begin{equation}
\tilde{Q}_{\rm Eff} = \cos{\left(\theta\right)}P + \sin{\left(\theta\right)}\tilde{Q}_{X_5^+} .
\end{equation}

The mixed mode transforms with same irreducible representation as the polarization in the cation ordered 10 atom cell. The strength of mixing is determined directly from the bilinear coupling coefficient $\tilde{\gamma}_2\left<\tilde{Q}_{X_3^-}\right>$ and the relative stiffnesses of the antipolar and polar modes $\Delta A = \frac{1}{2}\left(A_{X_5^+} - A_{P}\right)$, through the relation
\begin{equation}
\tilde{\gamma}_2\left<\tilde{Q}_{X_3^-}\right>\tan{\left(\theta\right)} = \Delta A - \sqrt{ \Delta A^{2} + \tilde{\gamma}_{2}^{2}\left<\tilde{Q}_{X_3^-}\right>^{2}} .
\end{equation}


\begin{thebibliography}{36}
\bibitem [{dis()}]{discrep}%
  \BibitemOpen
  \href@noop {} {}\bibinfo {note} {In the language of solid state
  chemistry{~\cite{brown09}}, one should maximize the difference in the A and
  A$^\prime$ discrepancy factors}\BibitemShut {NoStop}%
\end{thebibliography}
\end{document}